\documentclass[pra]{revtex4}
\usepackage{amsmath,amssymb,mathrsfs}
\usepackage{psfrag}
\usepackage{graphicx}
\usepackage{graphics}
\usepackage{epsfig}
\usepackage{bm}
\usepackage{color}
\usepackage{verbatim,color,ulem}

\newcommand{\beq}{\begin{equation}}
\newcommand{\eeq}{\end{equation}}
\newcommand{\beqa}{\begin{eqnarray}}
\newcommand{\eeqa}{\end{eqnarray}}
\newcommand{\ba}{\begin{aligned}[b]}
\newcommand{\ea}{\end{aligned}}

\begin{document}

\title{Interaction between gravitational waves and trapped
  Bose-Einstein condensates}
\author{A. Perodi$^{1}$ and L. Salasnich$^{1,2,3,4}$}
\affiliation{$^{1}$Dipartimento di Fisica e Astronomia "Galileo Galilei", 
Universit\`a di Padova, Via Marzolo 8, 35131 Padova, Italy
\\
$^{2}$Istituto Nazionale di Fisica Nucleare (INFN), Sezione di Padova, via Marzolo 8, 35131 Padova, Italy
\\
$^{3}$Padua QTech Center, Universit\`a di Padova, Via Gradenigo 6/A, 35131 Padova, Italy
\\
$^{4}$Istituto Nazionale di Ottica (INO) del Consiglio Nazionale delle Ricerche (CNR), via Nello Carrara 1, 50019 Sesto Fiorentino, Italy}

\begin{abstract}
  Inspired by recent proposals for detecting gravitational waves by using Bose-Einstein condensates (BECs), we investigate the interplay between these two phenomena. A  gravitational wave induces a phase shift in the fidelity amplitude of the many-body quantum state. We study the enhancement of the phase shift
  in the case of Bose condensates confined by an anisotropic harmonic
  potential, considering both ideal and interacting BEC.
\end{abstract}

\maketitle

\section{Introduction}

The connection between general relativity and Bose-Einstein condensates (BECs) is gaining increasing importance, both for the possibility of detecting events such as gravitational waves and for investigating still obscure aspects of Einstein's theory.
A recent development in this field is a theory known as emergent gravity, whose prime idea is that gravity is not a fundamental interaction and that spacetime is a composite object behaving approximately like a superfluid. It is thus possible to study certain cosmological phenomena in analogy with BECs \cite{roitberg2021einstein,roitberg2023emergence}, and to perform simulations of the effects of gravitational waves on them \cite{liberati2017analogue}. A further application of BECs on cosmological research is the possibility to exploit them for gravitational wave (GW) detection, in order to develop a more compact tool than today's interferometers (such as LIGO and Virgo \cite{abbott2016observation, abbott2017gw170814}) to investigate multimessanger astronomy. As shown in Ref. \cite{schutzhold2018interaction, sabin2014phonon, robbins2022detection}, the gravitational wave constitutes a source term for the phonons of the condensate and the intrinsic coherence of the latter provides an enhancement mechanism which could, at least in theory, compensate the extremely small magnitude of the gravitational wave which is of the order of $10^{-20}$. For a quite comprehensive review about the interaction between macroscopic quantum systems and gravity see, for instance, Ref. \cite{itastrange}. 

In this paper, we begin by deriving the gravitational wave (GW) equation and its solution in vacuum from the Einstein field equations in Sec. \ref{sec GW}. Subsequently, after providing a brief overview of the well-established description of a Bose-Einstein condensate (BEC) in flat space-time (see Sec. \ref{sec BEC}), we extend our analysis to a more general curved space-time and generalize the Gross-Pitaevskii equation in Sec. \ref{sec BEC curved}. Finally, in Sec. \ref{sec detectability}, we explore the interaction between GWs and BECs, assessing its detectability across various interaction strengths between the atoms constituting the condensate. To achieve this, it will be necessary to combine two theoretical frameworks: general relativity and quantum mechanics, which are complementary but still have few points of contact. For this purpose, the gravitational field will be considered classically as a geometric background on which all physical phenomena, described by quantum field theory (QFT), occur. Theoretically, an approach could involve quantizing the gravitational field and, consequently, treating both the gravitational and matter fields within a quantum-mechanical framework. {This approach has been pursued in various studies, including \cite{sen2024probing}, where the authors propose using BECs as a medium to detect gravitons originating from a stochastic GW background. While this line of inquiry is indeed intriguing, it lies beyond the scope of the present work.}

\section{Gravitational waves}
\label{sec GW}

Gravitational waves are perturbation of the metric of the spacetime and the equation that describes them is obtained starting from the Einstein field equations \cite{General_Relativity}
\begin{equation}
\label{eq campo einstein}
    R_{\mu\nu}-\frac{1}{2}R\,g_{\mu\nu}=\frac{8\pi G}{c^4}T_{\mu\nu}
\end{equation}
and imposing the weak field condition $g_{\mu\nu}=\eta_{\mu\nu}+h_{\mu\nu}$, 
where $\eta_{\mu\nu}=\text{diag}(-1,+1,+1,+1)$ is the Minkowski metric and $|h_{\mu\nu}|\ll1$. 
Proceeding with the calculations and choosing the Donder gauge $\partial^\mu h_{\mu\nu}-\frac{1}{2}\partial_\nu h_\alpha^\alpha=0$ and the TT-gauge (transverse-traceless gauge), which implies $h_\alpha^\alpha=0$, one gets the following result
\begin{equation}
\label{eq GW}
\left( {1\over c^2}\partial_t^2
-\nabla^2 \right)
h_{\mu\nu}=\frac{16\pi G}{c^4}T_{\mu\nu}
\end{equation}
which are indeed the equations of the GWs, with $\nabla^2=\partial_x^2+\partial_y^2+\partial_z^2$.  

Eq. (\ref{eq GW}) is not easily solvable analytically, primarily due to the expression of the energy-momentum tensor $T_{\mu\nu}$. Therefore, let's consider the simplest solution: in vacuum.
In this case the solution is a plane wave $ h^{\mu\nu}=\mathrm{Re}\left\{A^{\mu\nu}e^{ik_\sigma x^\sigma}\right\}$ where $k^\alpha k_\alpha=0$ and
\begin{equation}
\label{A TT}
    A^{\mu\nu}_{(TT)}=\left(\begin{array}{cccc}
         0&0&0&0\\
         0&A_+&A_\times&0\\
         0&A_\times&-A_+&0\\
         0&0&0&0
    \end{array}\right) \; . 
\end{equation}
From now on, we will only consider gravitational waves propagating along the $z-$axis with $+$ polarization, for which the metric perturbation takes the form $h^{\mu\nu}=\text{diag}(0,h,-h,0)$ 
which results in the modification of the infinitesimal line element
\begin{equation}
\label{ds^2}
    ds^2=g_{\mu\nu}dx^\mu dx^\nu=-c^2dt^2+(1+h)dx^2+(1-h)dy^2+dz^2 \; , 
\end{equation}
where $h=h(z,t)$ depends on $z$ and $t$.

\section{Bose-Einstein condensates}
\label{sec BEC}

The quantum-field-theory Hamiltonian for identical non-relativistic particles of mass $m$ in a flat spacetime is given by \cite{sala-book}
\begin{equation}
\hat{H}_{\rm flat}=\int d^3r\, \hat{\psi}^+(\vec r,t)\left[-\frac{\hslash^2}{2m}\nabla^2+U(\vec r)\right]\hat{\psi}(\vec r,t)
\nonumber
\end{equation}
\begin{equation}
+\frac{1}{2}\int d^3r\,d^3r'\,\hat{\psi}^+(\vec r,t)\hat{\psi}^+(\vec r',t)V(\vec r,\vec r')\hat{\psi}(\vec r',t)\hat{\psi}(\vec r,t) \; , 
\end{equation}
where ${\hat \psi}(\vec{r},t)$ is 
the quantum field operator, 
$U(\vec r)$ is the trapping potential, and $V(\vec r,\vec r')$ the inter-particle potential of the interaction between particles. In order to describe a bosonic field it is necessary to impose the following equal-time commutation rule
\begin{equation}
    [\hat{\psi}(\vec r,t),\hat{\psi}^+(\vec r',t)]=\delta(\vec r-\vec r') \; , 
\end{equation}
where $[{\hat A},{\hat B}]={\hat A}{\hat B}-{\hat B}{\hat A}$. 
When bosonic particles are cooled below the critical temperature $T_c$ (which depends on the particle density and is on the order of $\mathrm{nK}$) they lose their individuality and form a single entity where all particles occupy the same state. This phenomenon is known as a Bose-Einstein condensate, which is one of the few quantum phenomena observable on a macroscopic scale. 
In the case where $N\gg1$, with $N$ number of particles, and $T\ll T_c$, the field $\hat{\psi}(\vec r,t)$ acquires significant 'classical' relevance, and it is possible to exploit the many-body coherent states $|\Psi_{cs}\rangle$, introduced by Roy Glauber \cite{glauber1963coherent} and George Sudarshan \cite{sudarshan1963equivalence} in the '60, such that 
\begin{equation}
\hat{\psi}(\vec r,t)|\Psi_{cs}\rangle=\psi(\vec r,t)|\Psi_{cs}\rangle \; . 
\end{equation}
This method allows to study the condensate through the use of the complex bosonic scalar field $\psi(\vec r,t)$. In this way, we can circumvent the formalism of QFT and return to the more familiar theoretical framework of quantum mechanics by studying the wave function $\psi(\vec r,t)$, while keeping in mind that it represents a complex bosonic scalar field. 
For ultracold and dilue atoms one can safely impose a binary contact interaction 
\begin{equation}
V(\vec r,\vec r')=\gamma \, \delta(\vec r-\vec r') \; , 
\end{equation}
where $\gamma=4\pi\hslash^2 a_s/m$ with $a_s$ the scattering length. In this way one obtains 
the following Hamiltonian:
\begin{equation}
\label{H0 campo quantizzato}
    \hat{H}_{\rm flat}=\int d^3r\, 
    \left\{
    \hat{\psi}^+(\vec r,t)\left[-\frac{\hslash^2}{2m}\nabla^2+U(\vec r)\right]\hat{\psi}(\vec r,t)+\frac{\gamma}{2}\hat{\psi}^+(\vec r,t)^2\hat{\psi}(\vec r,t)^2
    \right\}
\end{equation}
from which, exploiting coherent states, the Gross-Pitaevskii equation (GPE) derives
\begin{equation}
\label{GPE 1}
    i\hslash\, \partial_t 
    \psi(\vec r,t)=\left[-\frac{\hslash^2}{2m}\nabla^2+U(\vec r)+\gamma|\psi(\vec r,t)|^2\right]\psi(\vec r,t) \; . 
\end{equation}
This equation, obtained independently by Eugene Paul Gross \cite{gross1961structure,gross1963hydrodynamics} and Lev Petrovi\u c Pitaevskii \cite{pitaevskii1961vortex}, is a nonlinear Schr\"odinger equation and it describes very accurately the BEC dynamics of the experiments with alkali-metal atoms. 

\section{Bose-Einstein condensates in curved spacetime}
\label{sec BEC curved}

In order to describe comprehensively the BECs in a curved spacetime we have to start from a Lorentz-invariant action, i.e. the Klein-Gordon one, to which the quartic term is added. Since the condensate is obtained at very low temperatures, therefore low thermal speeds, it is necessary to perform the non-relativistic limit. {By doing so one obtains the GPE in curved spacetime (see, for instance, Ref. \cite{schutzhold2018interaction,fagnocchi2010relativistic}), that corresponds to Eq. (\ref{GPE 1}) where the Laplace operator $\nabla^2$ is replaced by $\nabla_g^2$, which is the Laplace operator in curved spacetime.}

Imposing the arrival of a GW propagating in the $z-$direction with $+$ polarization (i.e. $g^{\mu\nu}=\eta^{\mu\nu}+h^{\mu\nu}$ with $h^{\mu\nu}=\text{diag}(0,h,-h,0)$) through the BEC one obtains the following equation
\begin{equation}
\label{GPE + h}
    i\hslash\, \partial_t \psi(\vec r,t)=\left[-\frac{\hslash^2}{2m}\left(\nabla^2+h(\partial_x^2-\partial_y^2)\right)+U(\vec r)+\gamma|\psi(\vec r,t)|^2 \right]\psi(\vec r,t) \; . 
\end{equation}
This is one of the main equations for our investigation of the interaction between GWs and BECs. {This mean-field equation is reliable provided that the nonlinear term is sufficiently small, namely for $\gamma \ll 1$.}

The detection of gravitational waves is a significant experimental milestone, precisely due to the magnitude of this phenomenon $h=\mathcal{O}(10^{-20})$. Interferometric techniques have made it possible to overcome this hurdle thanks to the use of resonant cavities, managing to gain the necessary orders of magnitude. 
The use of such large detectors, however, is not straightforward, and in view of a larger-scale implementation of multimessenger astronomy, other possibilities are being considered. We will therefore examine why BECs represents an opportunity in this regard.

\section{Quantum fidelity and enhanced phase shift}
\label{sec detectability}

Eq. (\ref{GPE + h}) suggests that the quantum-field-theory Hamiltonian of the bosonic scalar field interacting with a gravitational wave propagating along the z-axis with + polarization can be expressed as follows
\begin{equation}
    \hat{H}=\hat{H}_{\rm flat}+\hat{H}_{\rm int} \; , 
\end{equation}
where $\hat{H}_{\rm flat}$ is the Hamiltonian of the bosonic field in flat spacetime (Eq. (\ref{H0 campo quantizzato})), 
while $\hat{H}_{int}$ is defined as
\begin{equation}
\label{H int}
    \hat{H}_{\rm int}=\int \hat{\psi}^+(\vec r,t)\left(-\frac{\hslash^2}{2m}h(z,t)(\partial_x^2-\partial_y^2)  \right)\hat{\psi}(\vec r,t)\,d^3r \; .
\end{equation}

At this point, we can exploit a characteristic property of BECs, which are their coherent states $|\Psi_{cs}\rangle$ (of unit norm) which satisfy the following eigenvalue equation
\begin{equation}
\label{autovalori stati coerenti}
 \hat{\psi}(\vec r,t)|\Psi_{cs}\rangle=\psi(\vec r,t)|\Psi_{cs}\rangle=\sqrt{N}\, \phi_{gs}(\vec r) \, 
    e^{-{i\over\hbar}\mu t} 
    |\Psi_{cs}\rangle \; , 
\end{equation}
where the wavefunction $\psi(\vec r,t)$ is normalized to $N$, while the wavefunction $\phi_{gs}(\vec r)$, that is instead normalized to one, represents the ground-state of the BEC in flat spacetime with chemical potential $\mu$. Explicitlty, 
$\phi_{gs}({\vec r})$ satisfies the stationary GPE equation 
\begin{equation}
\left[-\frac{\hslash^2}{2m}\nabla^2+U(\vec r)+\gamma N |\phi_{gs}(\vec r)|^2 \right]\phi_{gs}(\vec r)
= \mu \, \phi_{gs}({\vec r}) \; . 
\end{equation}

We can now define the time evolution operator ${\hat U}_{\rm int}(t)$ associated with the interaction of the gravitational wave
\begin{equation}
    \hat{U}_{\rm int}(t)= \exp{\left(-\frac{i}{\hslash}\int_0^t\hat{H}_{\rm int}(t')\,dt'\right)} \; . 
\end{equation}
We also introduce the fidelity amplitude ${\cal F}(t)$, a complex number which quantifies of how much the many-body quantum state remains unchanged over time:
\begin{equation}
    {\cal F}(t)=
    \langle \Psi_{cs}|\hat{U}_{\rm int}(t)|\Psi_{cs}\rangle \; . 
\end{equation}
Notice that $F(t)=|{\cal F}(t)|^2$ is instead the familiar fidelity,
that is a non negative real number. 
Clearly, ${\cal F}(t)=1$ means no change while ${\cal F}(t)=0$ signals a
complete change, i.e. zero fidelity. 
From a theoretical point of view the gravitational wave will alter the
coherent state by shifting the value of $\digamma$ away from $1$ which is
desirable since quantum mechanics allows for the definite distinction of only
orthogonal states. By expanding the exponential in a Taylor series
in $Nh$, we obtain the following result:
\begin{equation}
\label{fidelity iNxi}
{\cal F}(t) 
=1-i N \, \xi(t) +
\mathcal{O}(N^2 h^2) \; , 
\end{equation}
where 
\begin{equation}
\label{def xi}
    N \, \xi(t)=N \frac{\hslash}{2m}\int_0^t dt'\int d^3r \, h(z,t')\phi_{gs}^\ast(\vec r)(\partial_y^2-\partial_x^2)\phi_{gs}(\vec r) \; . 
\end{equation}
    {Notice that the enhancement factor $N$ comes from the substitution
      reported in Eq. (\ref{autovalori stati coerenti}),
      since $\xi(t)$ depends only on the single-particle  time-independent wavefunction $\phi_{gs}(\vec r)$. 
We stress} that $\xi(t)$, being proportional to $h$, is of order $\xi=\mathcal{O}(10^{-20})$. However, as shown by Eq. (\ref{fidelity iNxi}), the time-dependent relative phase 
between the two many-body quantum states scales as $N \xi(t)$. The main reason to favor the detection of GWs through BECs is the possibility to make use of this enhancement mechanism proportional to the number of particles $N$ which, at least in theory, could compensate the small magnitude of $h$. 

Nonetheless, the experimental feasibility in creating condensate does not allow (to date where $N\approx 10^7$) to compensate over twenty orders of magnitude of the gravitational wave (i.e. $hN\approx 10^{-13})$, but new paths are opening wide towards the use of ultra cold Helium which consists in a lot more particles ($\mathcal{O}(10^{20})$), but on the other hand just less than $10\%$ of them are under the condensation threshold.

\subsection{Anisotropic harmonic confinement}
\label{sec Anisotropic harmonic confinement}

In this section we calculate $\xi(t)$, defined in Eq. (\ref{def xi}), for a Bose-Einstein condensate trapped by an anisotropic harmonic potential
\begin{equation}
\label{U arm anisotropo}
    U({\vec r})=\frac{1}{2}m(\omega_x^2x^2+\omega_y^2y^2+\omega_z^2z^2) \; . 
\end{equation}
By using the technique of  Feshbach resonances \cite{feshbach1958unified,feshbach1962unified,fano1935sullo} is now possible \cite{feshbach-now} to vary the interaction strength between the particles in a condensate, changing the scattering length $a_s$ even to the point of making it zero. 

\subsubsection{Non-interacting bosons}
\label{sec non-int}

For non-interacting bosons in the condensate, the ground-state wavefunction is given by 
\begin{equation}
\label{psi 1 gauss}    \phi_{gs}(\vec r)=\frac{1}{(\pi^3\sigma_x^2\sigma_y^2\sigma_z^2)^{\frac{1}{4}}} \, \exp{\left[-\left(\frac{x^2}{2\sigma_x^2}+\frac{y^2}{2\sigma_y^2}+\frac{z^2}{2\sigma_z^2}\right)\right]} \; , 
\end{equation}
where 
\begin{equation}
    \sigma_i=\sqrt{\frac{\hslash}{m\omega_i}} \quad \text{with} \; i=x,y,z
\end{equation}
The choice of a Gaussian shape trial function is well justified in the limit of weak interacting bosons, due to the fact that the exact ground-state of the linear Schr\"odinger equation with harmonic potential has this functional form. 
We can now insert Eq. (\ref{psi 1 gauss}) in Eq. (\ref{def xi}), obtaining
\begin{equation}
\begin{split}
    \xi(t)&=\frac{\hslash}{2m}\frac{\sigma_y-\sigma_x}{\sqrt{\pi}\sigma_x\sigma_y\sigma_z}\int_0^t dt'\int_{-\infty}^{+\infty}dz\;h(z,t')\;e^{-\frac{z^2}{\sigma_z^2}} \; . 
\end{split}
\end{equation}
Note that even in this case if the condensate is symmetric in the plane perpendicular to the propagation direction of the gravitational wave, i.e. $\sigma_x=\sigma_y$, the gravitational wave does not produce observable effects on it. Now using the simplest possible expression for the GW, which is 
\begin{equation}
\label{h z t esplicita}
    h(t,z)=h_0\cos(k_gz-\omega_g t)
\end{equation}
with $\omega_g=c k_g$, we find
\begin{equation}
\label{xi non int}
    N \xi(t)=N \frac{h_0}{2}\sqrt{\frac{\hslash}{m}}\left(\sqrt{\omega_x}-\sqrt{\omega_y}\right)\,e^{-\frac{\hslash}{4mc^2}\frac{\omega_g^2}{\omega_z}}\sin(\omega_gt) \; .
\end{equation}
This equation gives the phase shift, due to a GW, on a non-interacting BEC subjected to an anisotropic harmonic trapping potential. Note that the response of the condensate is oscillatory, as we would expect for a wave phenomenon.

\begin{figure}[h!]
    \includegraphics[scale=0.4]{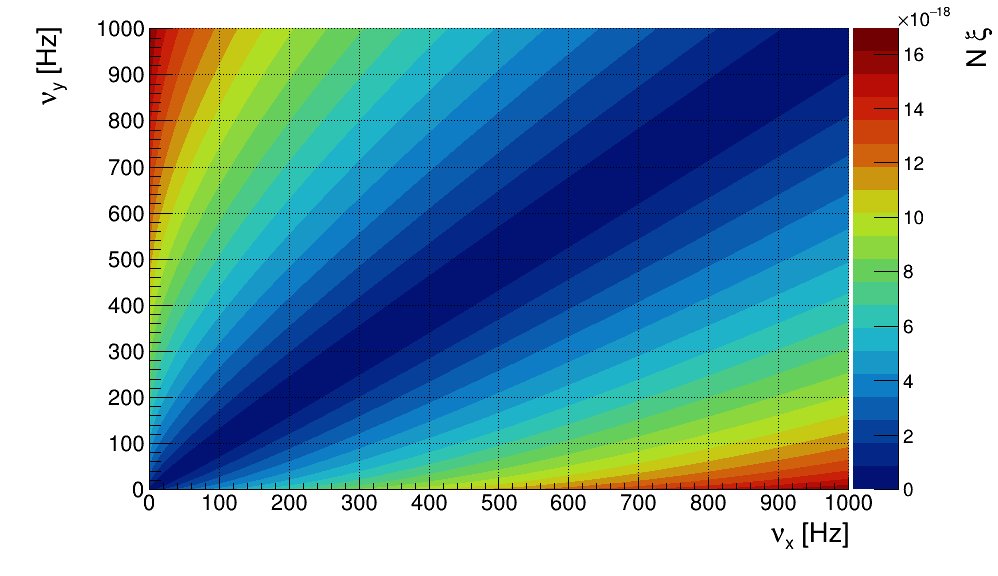}
    \caption{Phase shift $N \xi(t)$ at $t=\pi/(2\omega_g)$ as a function of the confinement frequencies in the $xy$ plane, for non-interacting Rubidium$-87$ atoms, with fixed values of $\nu_z=150\,\mathrm{Hz}$, $h_0=10^{-20}$ and $N=10^7$.}
    \label{fig:plot 2d}
\end{figure}

Fig. \ref{fig:plot 2d} shows a two-dimensional plot depicting the amplitude of $\xi(t)$, i.e. $N\xi(t=\pi/(2\omega_g))$,  
as a function of the confinement frequencies $\nu_x=\omega/(2\pi)$ and $\nu_y=\omega_y/(2\pi)$ 
in the $xy$ plane. The plot has been obtained considering a condensate consisting of non-interacting $^{87}$Rb atoms and keeping constant the values of $\omega_z=2\pi\times 150\,\mathrm{Hz}$, $h_0=10^{-20}$ and $N=10^7$; and shows how the phase amplitude induced by the GW and enhanced by the number of particles could currently reach the magnitude of $\mathcal{O}(10^{-17})$, if the confinement frequencies perpendicular to the propagation direction of the wave are appropriately chosen. This order of magnitude also confirms the approximation made by expanding the time evolution operator in a series. 
It is also important to stress that the amplitude of $N\xi$ is independent of both the gravitational wave frequency and the confinement frequency along the $z$ direction, since the exponential term in (\ref{xi non int}) is close to 1. Any significant damping effect on the amplitude would only be observable for values of $\omega_g^2 \gg 4mc^2\omega_z/\hbar$.

\subsubsection{Interacting bosons}
\label{sec weakly int}

The GPE is not easily solvable analytically, but nevertheless it is possible to make use of approximate methods to evaluate also the interacting case in which $\gamma\neq 0$. {Remind that the GPE is reliable if the nonlinear term is sufficiently small. In the case of harmonic confinement this condition means $Na_s/\ell_{H}\ll 1$, where  $\ell_H$ is the geometric mean of the characteristic harmonic lengths in the three directions: $\ell_H=(\ell_x \ell_y \ell_z)^{1/3}$ with $\ell_x=\sqrt{\hbar/(m \omega_X)}$, 
$\ell_y=\sqrt{\hbar/(m\omega_y)}$, 
and $\ell_z=\sqrt{m/(m\omega_z)}$.}
In \cite{salasnich2000time}, a variational approach was used with a Gaussian trial function of the same form of Eq. (\ref{psi 1 gauss}) 
whose $\sigma_i$ are variational parameters. {In other words, $\sigma_x$, $\sigma_y$, and $\sigma_z$ are not set equal to the three characteristic harmonic length $\ell_x$, $\ell_y$, and $\ell_z$. Instead, the three $\sigma_i$ are determined by minimizing the GPE energy functional.} 
In the cited article, it is shown that in the strongly repulsive regime the parameters $\sigma_i^\ast$ that minimize the ground-state energy have the following form:
\begin{equation}
\label{sigma var}
    \sigma_x^\ast=\left(\frac{\Gamma \hslash^{\frac{3}{5}}\omega_y\omega_z}{m\omega_x^4}\right)^{\frac{1}{5}} \; , 
    \quad
    \sigma_y^\ast=\left(\frac{\Gamma \hslash^{\frac{3}{5}}\omega_z\omega_x}{m\omega_y^4}\right)^{\frac{1}{5}} \; , 
    \quad
    \sigma_z^\ast=\left(\frac{\Gamma \hslash^{\frac{3}{5}}\omega_x\omega_y}{m\omega_z^4}\right)^{\frac{1}{5}} \; , 
\end{equation}
where 
\begin{equation}
    \Gamma = \gamma\frac{N}{(2\pi)^{\frac{3}{2}}} \; . 
\end{equation}
Given, as in Sec. \ref{sec non-int}, a Gaussian-type wave function, the procedure for calculating $N \xi(t)$ leads to to
\begin{equation}
\label{xi int var}
    N \xi(t)=N \frac{h_0}{2}\left(\frac{\hslash^{\frac{11}{5}}}{m^2\sqrt{\Gamma \omega_z}}\right)^{\frac{2}{5}}e^{-\frac{\omega_g^2}{4c^2}\left(\frac{\Gamma \hslash^{\frac{3}{5}}\omega_x\omega_y}{m\omega_z^4}\right)^{\frac{2}{5}}}\left(\left(\frac{\omega_x^4}{\omega_y}\right)^{\frac{1}{5}}-\left(\frac{\omega_y^4}{\omega_x}\right)^{\frac{1}{5}}\right)\sin(\omega_gt) \; . 
\end{equation}

\begin{figure}[h!]
    \includegraphics[scale=0.4]{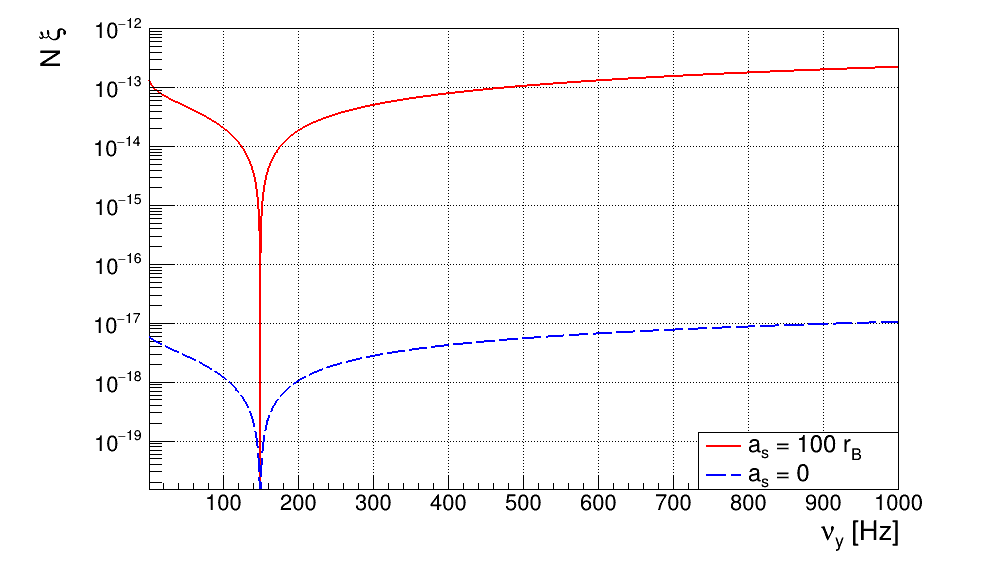}
    \caption{$N\xi(t)$ at $t=\pi/(2\omega_g)$ for a condensate of interacting Rubidium-87 atoms, plotted as a function of the confinement frequency along the $y$-axis, for various values of the scattering length $a_s$, with $r_B$ the Bohr radius. The parameters $\nu_z = 150\,\mathrm{Hz}$, $\nu_x = 150\,\mathrm{Hz}$, $h_0 = 10^{-20}$, and $N = 10^7$ were held constant.}
\label{fig:anda}
\end{figure}

Fig. \ref{fig:anda} shows the amplitude of $N\xi(t)$, for a condensate of interacting (red line) and non-interacting (blue line) $^{87}$Rb atoms, as a function of the 
confinement frequency along $y$. Assuming a value of $h_0=10^{-20}$ for the GW and $N=10^7$, the amplitude of the interacting BEC is of the order of $10^{-13}$, which in this case also confirms the approximation made by expanding the time evolution operator in a series. For typical values of the scattering length of $^{87}$Rb, i.e. around $100\,r_B$ (with $r_B$ the Bohr radius), the amplitude $N\xi$ is four orders grater than that related to the non-interacting BEC. This suggests that condensates of interacting atoms are preferable over the latter for future gravitational wave detection.

Note also that, as verified in the previous section, the amplitude is independent of the gravitational wave frequency, since the exponent of Eq. (\ref{xi int var}) is of the order of $\mathcal{O}(\omega_g^2\cdot10^{-36})$.

\section{Conclusions}

 We have discussed the interaction Hamiltonian between a gravitational wave and Bose-Einstein condensate, as reported in Eq. (\ref{H int}).
We have then computed the fidelity amplitude at first-order with respect to $hN$, obtaining Eq. (\ref{fidelity iNxi}), which shows an enhancement mechanism of the phase that is proportional to the number of particles constituting the condensate $N$.  Therefore, the latter could, at least in theory, serve as a potential mechanism for detecting gravitational waves.
Building upon the existing literature, we evaluate the magnitude of the interaction in the case of condensates confined by an anisotropic harmonic potential across three different configurations: non-interacting bosons, leading to Eq. (\ref{xi non int}) and interacting particles, resulting in Eq. (\ref{xi int var}). 
Additionally, as illustrated in Fig. \ref{fig:anda}, we have demonstrate that the effect of the gravitational wave is more pronounced for condensates of interacting atoms. 
As suggested by Sch\"utzhold, in the future one could experimentally produce a NOON-like state
\begin{equation}
|{\rm NOON}\rangle(0) = |\Psi_{cs}\rangle_{_A} \ |0\rangle_{_B} + |0\rangle_{_A} \
|\Psi_{cs}\rangle_{_B}  
\end{equation}
at time $t=0^-$, that is a macroscopic superposition of the
{ BEC} in two different configurations: $A$ and $B$. 
Then, if the { GW} interacts only with the { BEC} that is
in the configuration $A$, at time $t$ one has 
\begin{equation}
|{\rm NOON}\rangle(t) = e^{i N\xi(t)}
|\Psi_{cs}\rangle_{_A} \ |0\rangle_{_B} + |0\rangle_{_A} \ 
|\Psi_{cs}\rangle_{_B}  \; . 
\end{equation}
Thus, the NOON state acquires a relative phase that is exactly
the one of the fidelity amplitude, for a small time interval. NOON states
$|N\rangle_A|0\rangle_B+|0\rangle_A|N\rangle_B$ with up 
to $N = 10$ quanta were already produced with photons and phonons \cite{sandro1,sandro2}, however their
experimental realization with ultracold bosonic atoms is yet to be achieved. 

In conclusion, the main take-home message of our paper is that, at fixed number $N$ of atoms, tuning the s-wave scattering length $a_s$ of the inter-atomic interaction one can increase of several order of magnitude the many-body phase $N\xi(t)$ which signals the arrival of a gravitational wave. 


\section*{Authors contibution}

{Conceptualization, A.P. and L.S.; methodology, A.P. and L.S.; software, A.P.; writing-review and editing, A.P. and L.S.; visualization, A.P.; supervision, L.S.; funding acquisition, L.S.}

\section*{Acknowledgements}

{This research was funded by 
BIRD Project “Ultracold atoms in curved geometries” of the University of Padova; by the European Union-NextGenerationEU within the National Center for HPC, 13 Big Data and Quantum Computing [Project No. CN00000013, CN1 Spoke 10: Quantum Computing]; by the European
Quantum Flagship Project PASQuanS 2; by Iniziativa Speciﬁca Quantum of Istituto
Nazionale di Fisica Nucleare; by the Project Frontiere Quantistiche within the 2023 funding programme ’Dipartimenti di Eccellenza’ of the Italian Ministry for Universities and Research; by the PRIN 2022 Project Quantum Atomic Mixtures: Droplets, Topological Structures, and Vortices.}

\end{document}